\documentclass[iop]{emulateapj}
\usepackage{graphicx,amsmath,amssymb}
\usepackage{natbib}
\usepackage{apjfonts}
\usepackage{hyperref}
\bibpunct[, ]{(}{)}{;}{a}{}{,}

\begin{document}

   \title{A nearby GRB host prototype for $z\sim7$ Lyman-break galaxies:\\
          Spitzer-IRS and X-shooter spectroscopy of the host galaxy of
          GRB\,031203}

   \author{D.~Watson,\altaffilmark{1}
           J.~French,\altaffilmark{1}
           L.~Christensen,\altaffilmark{2,3}
           B.~O'Halloran,\altaffilmark{4}
           M.~Micha{\l}owski,\altaffilmark{5}
           J.~Hjorth,\altaffilmark{1}
           D.~Malesani,\altaffilmark{1}
           J.~P.~U.~Fynbo,\altaffilmark{1}
           K.~D.~Gordon,\altaffilmark{6}
           J.~M.~Castro~Cer\'on\altaffilmark{1,7}
           S.~Covino\altaffilmark{8}
           R.~F.~Reinfrank\altaffilmark{9,10}
            }
   \altaffiltext{1}{Dark Cosmology Centre, Niels Bohr Institute, University of Copenhagen, Juliane Maries Vej 30, DK-2100 Copenhagen \O, Denmark; darach, jens, malesani @dark-cosmology.dk}
   \altaffiltext{2}{Excellence Cluster Universe, Technische Universit\"{a}t M\"{u}nchen, Bolzmanstrasse 2, 85748 Garching, Germany}
   \altaffiltext{3}{European Southern Observatory, Karl-Schwarzschild-Strasse 2, 85748 Garching bei M\"{u}nchen, Germany}
   \altaffiltext{4}{Astrophysics Group, Imperial College, Blackett Laboratory, Prince Consort Road, London SW7 2AZ, UK}    
   \altaffiltext{5}{Scottish Universities Physics Alliance, Institute for Astronomy, University of Edinburgh, Royal Observatory, Edinburgh, EH9 3HJ, UK}
   \altaffiltext{6}{Space Telescope Science Institute, 3700 San Martin Drive, Baltimore, MD 21218, USA}
   \altaffiltext{7}{Herschel Science Centre (ESAC/ESA), Camino Bajo del Castillo, s/n, E-28.692 Villanueva de la Ca\~{n}ada (Madrid), Spain}
   \altaffiltext{8}{INAF/Osservatorio Astronomico di Brera, via Emilio Bianchi 46, 23807 Merate (LC), Italy}
   \altaffiltext{9}{Australia Telescope National Facility, CSIRO, P.O. Box 76, Epping, NSW 1710, Australia}
   \altaffiltext{10}{School of Chemistry \& Physics, The University of Adelaide, Adelaide, SA 5005, Australia}

   \begin{abstract}
Gamma-ray burst (GRB) host galaxies have been studied extensively in optical
photometry and spectroscopy.  Here we present the first mid-infrared
spectrum of a GRB host, HG\,031203.  It is one of the nearest GRB hosts at
$z=0.1055$, allowing both low and high-resolution spectroscopy with
\emph{Spitzer}-IRS.  Medium resolution UV-to-$K$-band spectroscopy with the
X-shooter spectrograph on the VLT is also presented, along with
\emph{Spitzer} IRAC and MIPS photometry, as well as radio and sub-mm
observations.  These data allow us to construct a UV-to-radio spectral
energy distribution with almost complete spectroscopic coverage from
0.3--35\,$\mu$m of a GRB host galaxy for the first time, potentially
valuable as a template for future model comparisons.  The IRS spectra show
strong, high-ionisation fine structure line emission indicative of a hard
radiation field in the galaxy -- in particular the [\ion{S}{4}]/[\ion{S}{3}]
and [\ion{Ne}{3}]/[\ion{Ne}{2}] ratios -- suggestive of strong ongoing
star-formation and a very young stellar population.  The absence of any PAH
emission supports these conclusions, as does the probable hot peak dust
temperature, making HG\,031203 similar to the prototypical blue compact
dwarf (BCD) galaxy, II\,Zw\,40.  The selection of HG\,031203 via the
presence of a GRB suggests that it might be a useful analogue of very young
star-forming galaxies in the early universe, and hints that local BCDs may
be used as more reliable analogues of star-formation in the early universe
than typical local starbursts.  We look at the current debate on the ages of
the dominant stellar populations in $z\sim7$ and $z\sim8$ galaxies in this
context.  The nebular line emission is so strong in HG\,031203, that at
$z\sim7$, it can reproduce the spectral energy distributions of $z$-band
dropout galaxies with elevated IRAC $3.6$ and $4.5\,\mu$m fluxes without the
need to invoke a $4000$\,\AA\ break.  Indeed, photometry of HG\,031203 shows
elevation of the broadband $V$-magnitude at a level similar to the IRAC
elevation in stacked $z$-band dropouts, solely due to its strong
[\ion{O}{3}] line emission.
   \end{abstract}
   \keywords{ gamma-ray burst: general -- gamma-ray burst: individual:
              GRB031203 -- early universe -- dark ages, reionisation, first
              stars -- galaxies: dwarf -- galaxies: ISM
             }

   \maketitle

%
%
\section{Introduction\label{introduction}}

Long-duration gamma-ray bursts (GRBs) trace the formation of massive stars
\citep{1998Natur.395..670G,2003Natur.423..847H,2003ApJ...591L..17S,2002AJ....123.1111B}. 
They occur across a vast range of cosmic distances from $z=0.0085$
\citep{1998Natur.395..670G} to $z=8.2$ \citep{2009Natur.461.1254T}, and with a very high
mean redshift \citep[$\langle z\rangle=2.2$][]{2009ApJS..185..526F}.  Their hosts
therefore provide a unique way to select star-forming galaxies because,
unlike other selection techniques, their selection is unrelated to the
emission properties of the galaxies themselves.

Studies of GRB hosts have so far suggested that they are star-forming
galaxies with relatively high specific star-formation rates and young
stellar populations
\citep{2004A&A...425..913C,2006ApJ...653L..85C,2009ApJ...691..182S,2010MNRAS.405...57S,2002ApJ...566..229C}. 
However, it is worth bearing in mind that such general statements are still
difficult to make in an absolute sense because of the potential bias
introduced in sample selection because most GRB hosts have been selected up
to now on the basis of localisations dependent on optical afterglows.  Some
attempts to address this essential question are being made with
X-ray--selected samples \citep{2009AIPC.1111..513M}.

Optical spectroscopy in absorption using the afterglow
\citep[e.g.][]{2009ApJS..185..526F}, or occasionally in emission from the
host galaxy itself where it is bright enough
\citep[e.g.][]{2007A&A...464..529W,2010AJ....139..694L} have indicated
metallicities in GRB host galaxies that are relatively low, but not atypical
of the population forming high-mass stars, at least at $z\gtrsim2$
\citep{2006A&A...451L..47F,2009ApJ...693.1236C}.  One of the best-studied
cases so far, because of its proximity at $z=0.1055$, is the host of
GRB\,031203 \citep[HG\,031203,][]{2004ApJ...611..200P}.  These studies
\citep{2004ApJ...611..200P,2007A&A...474..815M,2010AJ....139..694L,2010A&A...514A..24H}
have shown HG\,031203 to be a relatively low metallicity system with a
fairly high star-formation rate and a young stellar population.  As with all
optical studies, such analyses are dominated by the low-dust regions and may
give an incomplete impression of the galaxy since a significant fraction of
the stellar emission, particularly from young stars, may be re-processed by
dust and emerge in the mid- and far-infrared (IR).

To date, all spectroscopic studies of GRB host galaxies have been at optical
or near-IR wavelengths. Here we present the first spectroscopic
study of a GRB host galaxy in the mid-IR
($5$--$40\,\mu$m) using the \emph{Spitzer Space Telescope} \citep{2004ApJS..154....1W} Infrared
Spectrograph \citep[IRS][]{2004ApJS..154...18H}. We also present an almost continuous spectral energy
distribution (SED) for this galaxy from the ultraviolet through to the
mid-IR ($0.35$--$40\,\mu$m), with detections and upper limits at far-IR,
sub-mm and radio wavelengths, the first such SED for a GRB host.

Uncertainties quoted are at the 68\% confidence level for one interesting
parameter unless otherwise stated. A cosmology where
$H_0=72$\,km\,s$^{-1}$\,Mpc$^{-1}$, $\Omega_\Lambda = 0.73$ and $\Omega_{\rm
m}=0.27$ is assumed throughout.

%
%
\section{Observations and data reduction}\label{observations} 

\subsection{X-shooter}

HG\,031203 was observed during the commissioning of the second generation
VLT instrument, X-shooter \citep{2006SPIE.6269E..98D} on 17 March 2009. 
X-shooter is a single object echelle spectrograph consisting of three arms
which cover simultaneously the spectral range 300--2480\,nm.  The slit
widths chosen during the observations were 1\farcs0 in the UVB arm
($\Delta\lambda = 300$--550\,nm), and 0\farcs9 in the VIS ($\Delta\lambda =
550$--1015\,nm) and NIR arms ($\Delta\lambda = 1000$--2480\,nm).  This
instrument setup gives resolutions in the three arms of $R=5100$, 8800, and
5600, respectively.  The slit width matched the seeing during the
observation.  During the observations the slit was aligned along the
parallactic angle.  We obtained four 20 minute integrations on target, with
offsets along the slit in a classic ABBA observing pattern.  The UVB
detector was binned by a factor of two in the spectral direction when the
detectors were read out.  With a 1\arcsec\ slit width in the UVB arm, the
sampling for the full-width half-maximum of a line is 6 pixels, so a better
signal-to-noise ratio per pixel is obtained in binned data.

Data reduction was performed with a preliminary version of the
pipeline \citep{2006SPIE.6269E..80G}. Tracing the individual
orders, wavelength calibration, and flat fielding were done using
calibration frames obtained during the commissioning run. In the UVB
and VIS data, subtraction of the sky background used a routine
described in \citet{2003PASP..115..688K}, while the background in
the NIR data was subtracted using the adjacent science
exposures. Finally the individual orders were extracted and merged
using a weighted mean combination scheme, resulting in a
two-dimensional spectrum for each arm. Similar data reduction
procedures were done for the HST white dwarf spectrophotometric
standard star GD\,71 which was used to flux calibrate the science data,
and for an observation of the O8V star Hipparcos\,69892, which provided
a reference smooth spectrum for dividing out telluric absorption lines
from the earth's atmosphere.

Further analysis of the data was performed with conventional IRAF
routines. The four two-dimensional spectra were combined, and
one-dimensional spectra were extracted using an aperture which matched
the three arms. In the flux calibration of the observation, we
included a correction for the atmospheric extinction using an
extinction curve appropriate for Paranal.

\subsection{\emph{Spitzer}}

The \emph{Spitzer} photometric data on HG\,031203 were obtained with IRAC
\citep{2004ApJS..154...10F} at 3.6 and $5.8\,\mu$m with a 30\,s frame time,
and a total integration time of 360\,s and MIPS \citep{2004ApJS..154...25R}
at $24\,\mu$m also with 30\,s integration times and a total observation time
of $\sim1350$\,s.  The data, under program 20370, were reduced as outlined
in \citet{2010ApJ...721.1919C}.

The target was also observed spectroscopically, in staring mode using all
four IRS modules; short-low (SL, 5.2--14.5\,$\mu$m), long-low (LL,
14.0--38.0 \,$\mu$m), short-high (SH, 9.9--19.6\,$\mu$m), and long-high (LH,
18.7--37.2\,$\mu$m).  The resolution of the low-resolution modules is
$\sim$60--127, while that of the high-resolution modules is $\sim$600. 
Observations were carried out on 28 May 2005 (see
Table~\ref{tab:observations}).

\begin{table}
\caption{Spitzer-IRS observations of the host galaxy of GRB\,031203\label{tab:observations}}
\begin{tabular}{@{}lllllll@{}}
\hline\hline 
Module & SH & LH & SL1 & SL2 & LL1 & LL2 \\
\hline
Integration Time (s) & 3656.9 & 1828.5 & 609.5 & 1828.5 & 314.6 & 943.7 \\ 
\hline
\end{tabular}
\end{table}

Data from the short-low and short-high modules were preprocessed with the
\textit{Spitzer} Science Centre (SSC) data reduction pipeline version 15.3,
while data from the long-low and long-high modules were processed with
version 17.2.

For the high-resolution modules, spectra were extracted from each individual
basic calibrated data (BCD) file using the Spitzer IRS Custom Extraction
(SPICE) software. Prior to extraction of the spectra, data files were
individually cleaned of rogue pixels using the IDL program IRSCLEAN version
1.9. The rogue pixel masks used for cleaning were a combination of the
default campaign masks provided by the SSC and masks generated automatically
by running IRSCLEAN's rogue pixel identification algorithm.

Individual spectra were subsequently combined with a clipped mean algorithm
using the Spectral Modelling, Analysis, and Reduction Tool (SMART) developed
by the IRS team \citep{2004PASP..116..975H}.  The edges of each order were
trimmed based on the wavelength calibration ranges provided in Table 5.1 of
the IRS Data Handbook version 3.1.  Since no sky measurements were taken,
the contribution of the sky emission was not subtracted from the
high-resolution spectra.  No subtracting the sky background has a
significant effect on the level of the continuum measured with the
high-resolution modules.  However, we use the high-resolution data only to
extract emission line fluxes, not for the SED work, so that a continuum
offset is not a significant issue.  The fluxes obtained with the
high-resolution modules are consistent with those retrieved from the low
resolution spectra.

For the low-resolution data, we made use of pipeline co-added,
sky-subtracted products. These files result from co-adding the
data from each nod position, then subtracting each nod position from the
other in order to subtract the background. The spectra were cleaned for
rogue pixels and optimally extracted to produce the final 1-D spectra.

The source was also observed with MIPS in SED mode. The data were reduced in
a standard way, as described in \citet{2010AJ....139.1553V}.

\begin{figure*}
 \includegraphics[width=\textwidth,viewport=8 8 683 337,clip=]{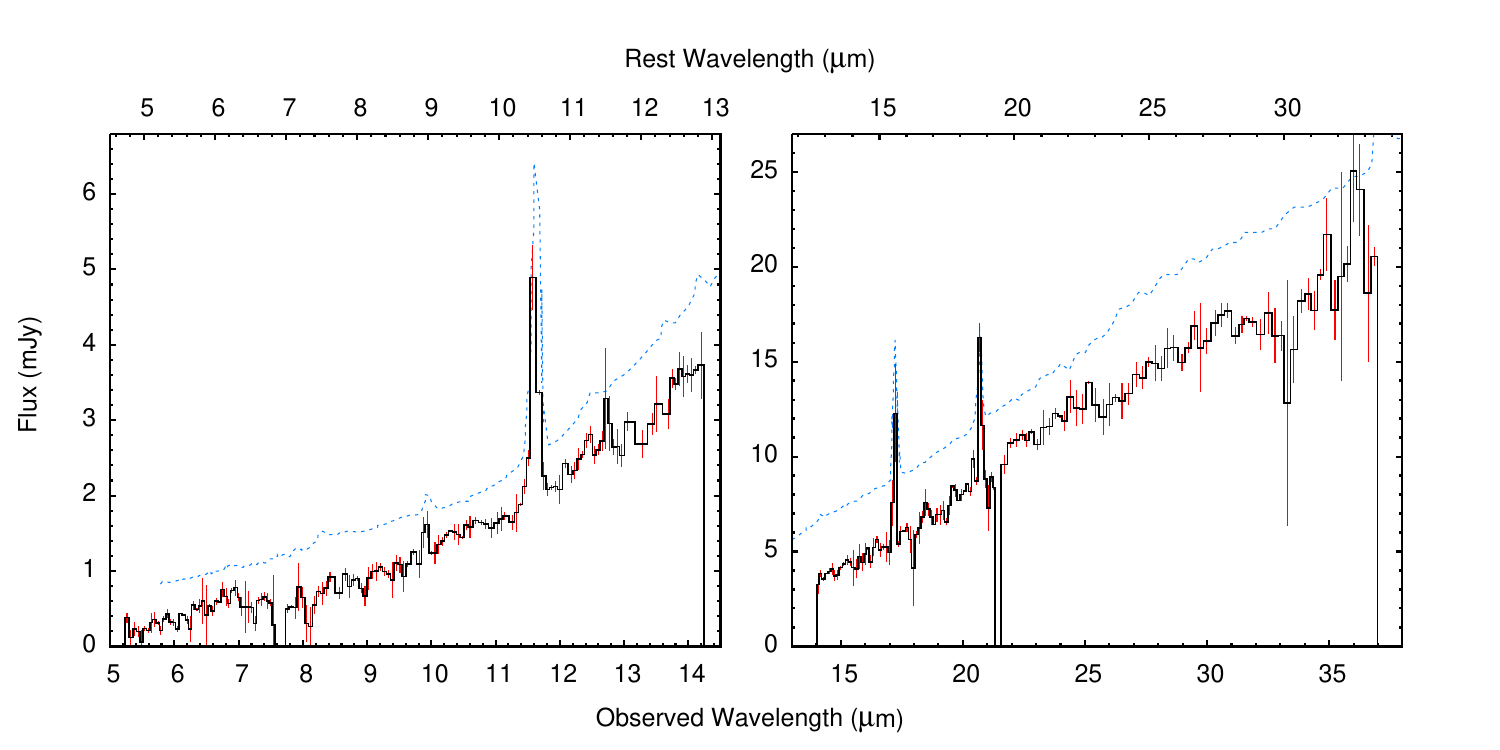}
\caption{Low-resolution IRS spectrum of the host galaxy of GRB\,031203.
 Orders 1 and 2 of the SL (left) and LL (right) modules. Thin red lines
 indicate uncertainties on the spectrum. Strong fine
 structure lines typical of a young star-forming galaxy are detected (see
 Fig.~\ref{fig:lines} for close-ups of the lines as observed with the
 high-resolution modules).  Very high-ionisation fine structure lines,
 indicative of AGN activity \protect \citep[e.g.\
 {[}\ion{Ne}{5}{]}\,$14.3\,\mu$m or
 {[}\ion{O}{4}{]}\,$25.89\,\mu$m,][]{2010ApJ...709.1257T}, are not found.
 The IRS spectrum of the blue compact dwarf galaxy II\,Zw\,40 from
 \protect\citet{2006ApJ...639..157W}
 is plotted (dashed line), scaled and offset, for comparison purposes.}
 \label{fig:IRSspec}
\end{figure*}

\subsection{Sub-mm--radio}

We obtained radio continuum observations of HG\,031203 on 26 January 2008
using the Australia Telescope Compact Array (ATCA) in configuration 6A, with antennae positioned on both
east-west and north-south tracks, and baselines of 60--4500\,m. 
Simultaneous observations were made at 20\,cm (1.39\,GHz) and 13\,cm
(2.37\,GHz), with a bandwidth of 128\,MHz at each frequency.  A total of
7\,hr of on-source data were obtained.  Calibrator sources PKS\,B1934$-$638
and PKS\,B0826$-$373 were utilized to set the absolute flux calibration of
the array and to calibrate phases and gains, respectively.  Data reduction
and analysis was done using the MIRIAD package \citep{miriad}.  The final
synthesized beam sizes for 20 and 13\,cm images were
$8\farcs5\times3\farcs4$ and
$6\farcs3\times2\farcs3$, respectively, with root-mean-square (rms) values of $46$
and $37\,\mu$Jy\,beam$^{-1}$.  The flux density of HG\,031203 was
estimated by fitting a two-dimensional Gaussian function to the data with
its centroid, size, and orientation as free parameters.

We obtained submillimeter ($870\,\mu$m) observations on 13--18 August 2008
using the Large Apex BOlometer CAmera \citep[LABOCA;][]{2009A&A...497..945S}
mounted at the Atacama Pathfinder Experiment
\citep[APEX;][]{2006A&A...454L..13G}.  A total of 7\,hr of on-source data
were obtained.  The weather condition varied between $0.3$--$1.4$\,mm of
precipitable water vapour.  Data reduction and analysis was done using the
miniCRUSH package
\citep{2008SPIE.7020E..45K}\footnote{\href{http://www.submm.caltech.edu/~sharc/crush/}{\texttt{http://www.submm.caltech.edu/\textasciitilde sharc/crush/}}}. 
We used the `deep' option that results in the best signal-to-noise ratio for
faint, point-like objects.  The beam size for the final image was
$19\farcs5$ and the rms was $2.2$\,mJy\,beam$^{-1}$.  The target was not
detected.

%
%
\section{Results}\label{results}

\emph{Spitzer}, APEX and ATCA each provide photometry or photometric limits
at long wavelengths, allowing us to determine the extinction-free emission
from HG\,031203.  These data are presented in Table~\ref{tab:photometry}. 
The \emph{Spitzer} photometric detections are consistent with the
flux-calibrated IRS spectra at the same wavelengths.

\begin{table}
\caption{Spitzer, APEX, and ATCA photometry of HG\,031203}
\label{tab:photometry}
\setlength{\tabcolsep}{3pt}
 \begin{center}
  \begin{tabular}{@{}lccccccc@{}}
   \hline\hline
Instr. &   \multicolumn{2}{c}{IRAC} & \multicolumn{2}{c}{MIPS} & APEX & \multicolumn{2}{c}{ATCA}\\
       &   \multicolumn{2}{c}{($\mu$Jy)} & \multicolumn{2}{c}{(mJy)} & (mJy) & \multicolumn{2}{c}{($\mu$Jy)}\\
Band &    $3.6\,\mu$m & $5.8\,\mu$m & $24\,\mu$m & SED & $870\,\mu$m & 13\,cm & 21\,cm \\
   \hline
Flux &    $193\pm2$ & $356\pm13$ & $11.3\pm0.4$ & $<40$ & $<12$ & $191\pm37$ & $254\pm46$\\
   \hline
  \end{tabular}
 \end{center}
\end{table}

\subsection{Spitzer spectroscopy}

\begin{figure}
 \includegraphics[angle=0,width=\columnwidth]{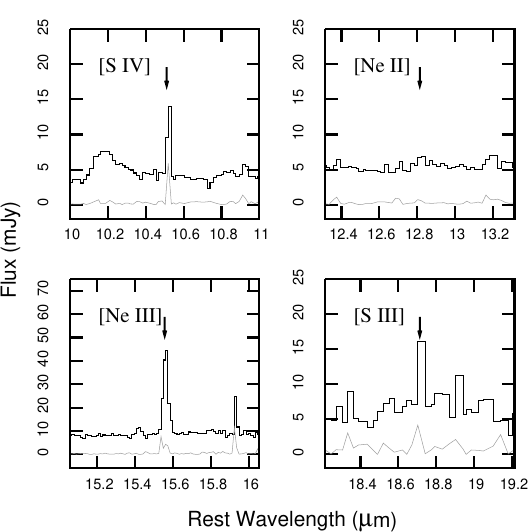}
 \caption{The locations of the most prominent mid-IR fine structure lines
          typically found in star-forming galaxies from the high-resolution,
          SH and LH, modules.  The grey lines indicate the associated
          uncertainty at each wavelength.  The broad feature at $10.2\,\mu$m
          is undetected in the SL spectrum and is related to an unremoved
          artefact in our SH data.}
 \label{fig:lines}
\end{figure}

The most striking features of the IRS spectra of HG\,031203
(Fig.~\ref{fig:IRSspec}) are clearly the strong forbidden lines,
[\ion{S}{4}]\,$10.51\,\mu$m, [\ion{S}{3}]\,$18.71\,\mu$m,
[\ion{Ne}{3}]\,$15.55\,\mu$m, and [\ion{Ne}{2}]\,$12.81\,\mu$m
(Fig.~\ref{fig:lines} and Table~\ref{tab:emissionlines}).

\begin{table}
\caption{Principal emission lines in the mid-IR spectrum of HG\,031203}
\label{tab:emissionlines}
\setlength{\tabcolsep}{6pt}
 \begin{center}
  \begin{tabular}{@{}lll@{}}
   \hline\hline
   Line ID		& Wavelength ($\mu$m)	& Flux ($10^{-15}$\,erg\,cm\,s$^{-1}$)	\\
   \hline
   {[}\ion{S}{4}{]}	& 10.51 		& $8.5\pm1.5$				\\
   {[\ion{Ne}{2}]} 	& 12.81 		& $0.7\pm0.2$				\\
   {[\ion{Ne}{3}]} 	& 15.56 		& $10.6\pm1.1$				\\
   {[\ion{S}{3}]} 	& 18.71 		& $5.2\pm0.7$				\\
   \hline
  \end{tabular}
 \end{center}
\end{table}

Comparing our
spectrum to spectra of other star-forming galaxies
\citep{2006ApJ...639..157W}, it is immediately apparent that the broad
features at 6.2, 7.7, 8.6 and 11.2, 12.7, and $16.4\,\mu$m, believed to be
associated with polycyclic aromatic hydrocarbons
\citep[PAHs,][]{2008ARA&A..46..289T}, are, somewhat surprisingly, not
detected.  The lack of PAH emission could be indicative of a very strong
continuum from an active galactic nucleus (AGN) diluting the PAH features, a
discriminant noticed with ISO
\citep[e.g.][]{1998ApJ...505L.103L,2000A&A...359..900L}.  The presence of an
AGN is claimed by \citet{2010AJ....139..694L}, however, even at first glance
this seems unlikely since the PAH emission appears to be simply absent,
rather than diluted by a powerful continuum.  We address the other lines of
evidence that essentially rule out a significant AGN contribution in this
galaxy below.  The other possibility is that the galaxy is a powerful
starburst.  In such galaxies, it is well-known that the PAH flux is
anti-correlated with the hardness of the radiation field, either because
young star-forming populations do not form PAHs, or because the hard
radiation field destroys them \citep{2006ApJ...639..157W}.  This seems
consistent with our findings from the mid-IR and optical--near-IR spectra
below.

\begin{figure}
 \includegraphics[angle=0,viewport=20 5 337 317,clip=,width=\columnwidth]{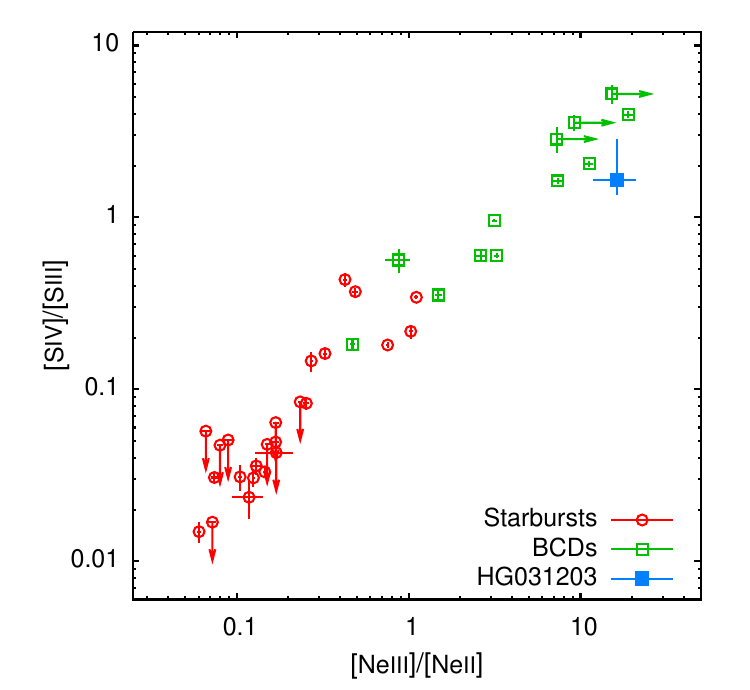}
 \caption{Line ratios of star-forming galaxies. The starburst galaxy sample
          of \protect\citet{2009ApJS..184..230B} are plotted as circles. A
          sample of blue compact dwarf galaxies, which have younger dominant
          stellar populations, higher specific star-formation rates, and
          harder radiation fields even than the starburst galaxies, is also
          plotted \protect\citep[open squares,][]{2006ApJ...639..157W}. 
          HG\,031203 (filled square) is among the most extreme of the blue
          compact dwarf galaxies, indicating that its dominant stellar
          population is extremely young.  The fact that the diagnostic lines
          are in the mid-IR ensures that the diagnostics are not
          significantly affected by dust obscuration.}
 \label{fig:line_ratios}
\end{figure}

The ratios of the forbidden lines, associated with the excitation state of
the gas in the interstellar medium of the galaxy, indicate the hardness of
the radiation field.  In Fig.~\ref{fig:line_ratios} we show the
[\ion{S}{4}]/[\ion{S}{3}] flux ratio ($1.6_{-0.3}^{+1.2}$) as a function of
[\ion{Ne}{3}]/[\ion{Ne}{2}] ($16.3\pm4.5$), and compare HG\,031203 to
samples of starburst galaxies and blue compact dwarf galaxies (BCDs).  The spectrum shows
no indication of absorption at $9.7\,\mu$m from silicates, neither does it
show any hint of the higher excitation lines that might be expected from an
AGN such as [\ion{Ne}{5}]\,$14.3\,\mu$m or [\ion{O}{4}]\,$25.9\,\mu$m
\citep{2010ApJ...709.1257T}.  All of these facts point to HG\,031203 being a
purely star-forming galaxy, but with an exceptionally hard radiation field,
indicating a very young stellar population, probably $\lesssim10$\,Myr
\citep{2000NewAR..44..249M}.  The fact that these ratios are
observed in the MIR suggests that the galaxy has essentially the same
character throughout, i.e.\ that it is dominated in the MIR by the same
young stellar population that dominates in the optical (see below).

\subsection{X-shooter spectroscopy}

\begin{figure*}
 \begin{center}
  \includegraphics[angle=0,viewport=74 62 746 576,clip=,width=0.85\textwidth]{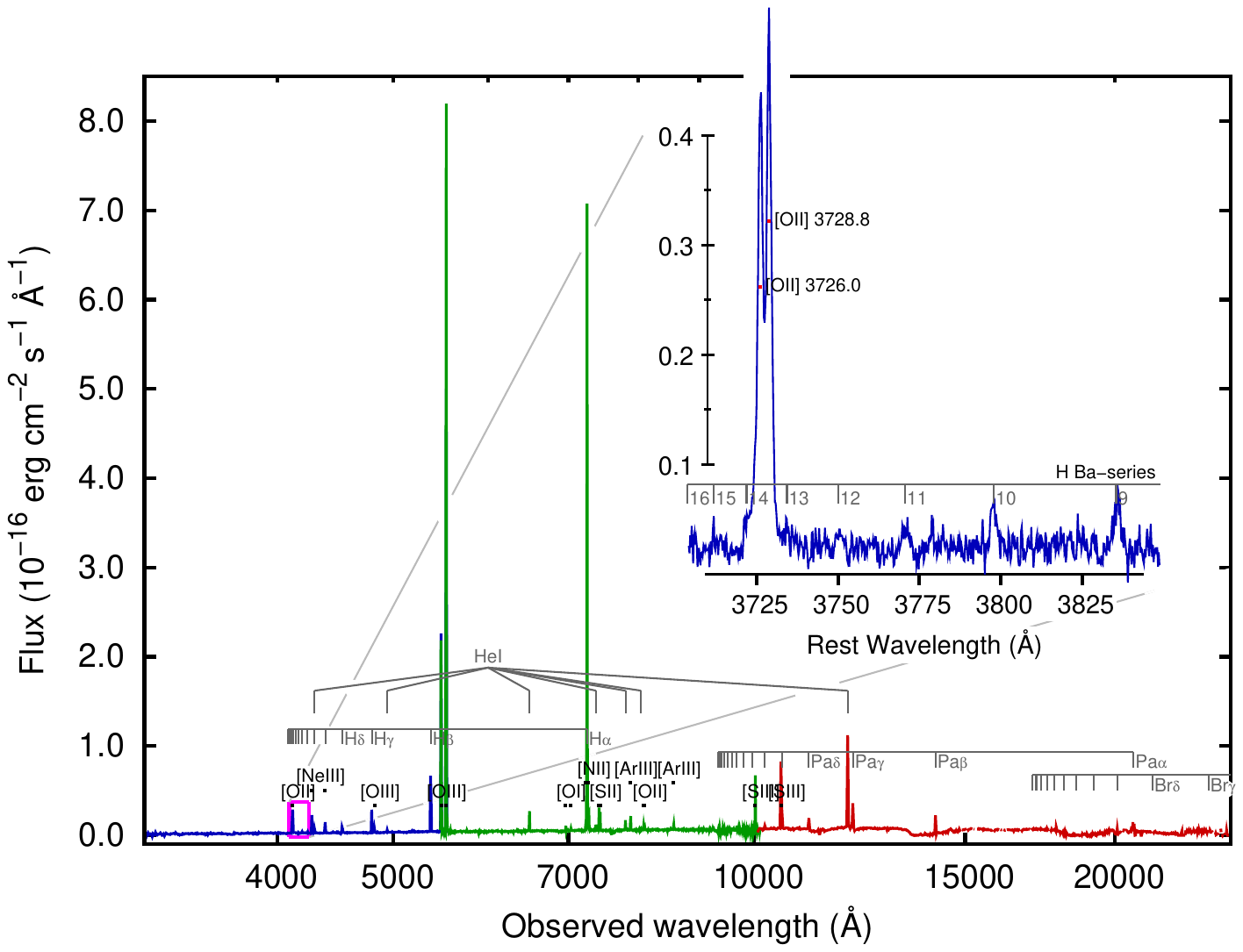}
  \caption{The X-shooter spectrum of HG\,031203. Strong nebular emission
           dominates the spectrum.
           }
 \label{fig:X-shooterspec}
 \end{center}
\end{figure*}

The X-shooter spectra are very information-rich
(Fig.~\ref{fig:X-shooterspec}), containing many strong emission lines with a
resolving power high enough to look in some detail at the interstellar
medium (ISM) in this galaxy.  However in this paper we will address
ourselves primarily to the continuum, with regard to the analysis of the SED
of the galaxy in order to determine the dominant emission processes and the
overall nature of the ISM.  In this regard, we have also determined the
ratios of the fluxes of nebular lines, to determine the nature of the source
powering the optical emission: [\ion{O}{2}], H$\beta$, [\ion{O}{3}],
[\ion{S}{2}], [\ion{N}{2}], and [\ion{O}{1}].  The relevant line ratios are:
log([\ion{O}{3}]/H$\beta) = 0.77\pm0.02$, log([\ion{S}{2}]/H$\alpha) =
-1.23\pm0.02$, log([\ion{N}{2}]/H$\alpha) = -1.35\pm0.02$,
log([\ion{O}{1}]/H$\alpha) =
-2.02^{+0.06}_{-0.07}$, and log([\ion{O}{3}]/[\ion{O}{2}]$) =
0.78^{+0.06}_{-0.07}$ .

These high-quality ratios resolve the debate about the presence of an AGN in
HG\,031203.  While the line ratios do indicate a very hard radiation field
in the ISM of this galaxy, making it among the most extreme starburst
galaxies consistent with the MIR line ratios, they unequivocally locate it
in the star-forming region of the Baldwin-Phillips-Terlevich
\citep[BPT,][]{1981PASP...93....5B} excitation diagrams \citep[see][and
references therein]{2006MNRAS.372..961K}, contrary to a recent claim
\citep{2010AJ....139..694L}, and consistent with previous findings of a pure
starburst
\citep{2004ApJ...611..200P,2007A&A...474..815M,2010A&A...514A..24H}.  Our
similar finding from the mid-IR spectrum above on the absence of AGN
characteristics, indicates that there is no very different obscured
component.

The presence of a break in the spectrum at 4000\,\AA\ is an indication of
the age of the stellar population \citep{2003MNRAS.344.1000B}.  We
quantified the break in HG\,031203 using the definition of D4000 of
\citet{1999ApJ...527...54B}, finding D4000~=~$0.93\pm0.05$.  This
effectively means that there is no 4000\,\AA\ break, and indicates a
continuum dominated by a very young stellar population, probably younger
than $\sim10$\,Myr, with a high specific star-formation rate
\citep{2004MNRAS.351.1151B,2005MNRAS.362...41G}.

\subsection{The spectral energy distribution of HG\,031203}

\begin{figure*}
 \includegraphics[angle=0,viewport=50 149 744 471,clip=,width=\textwidth]{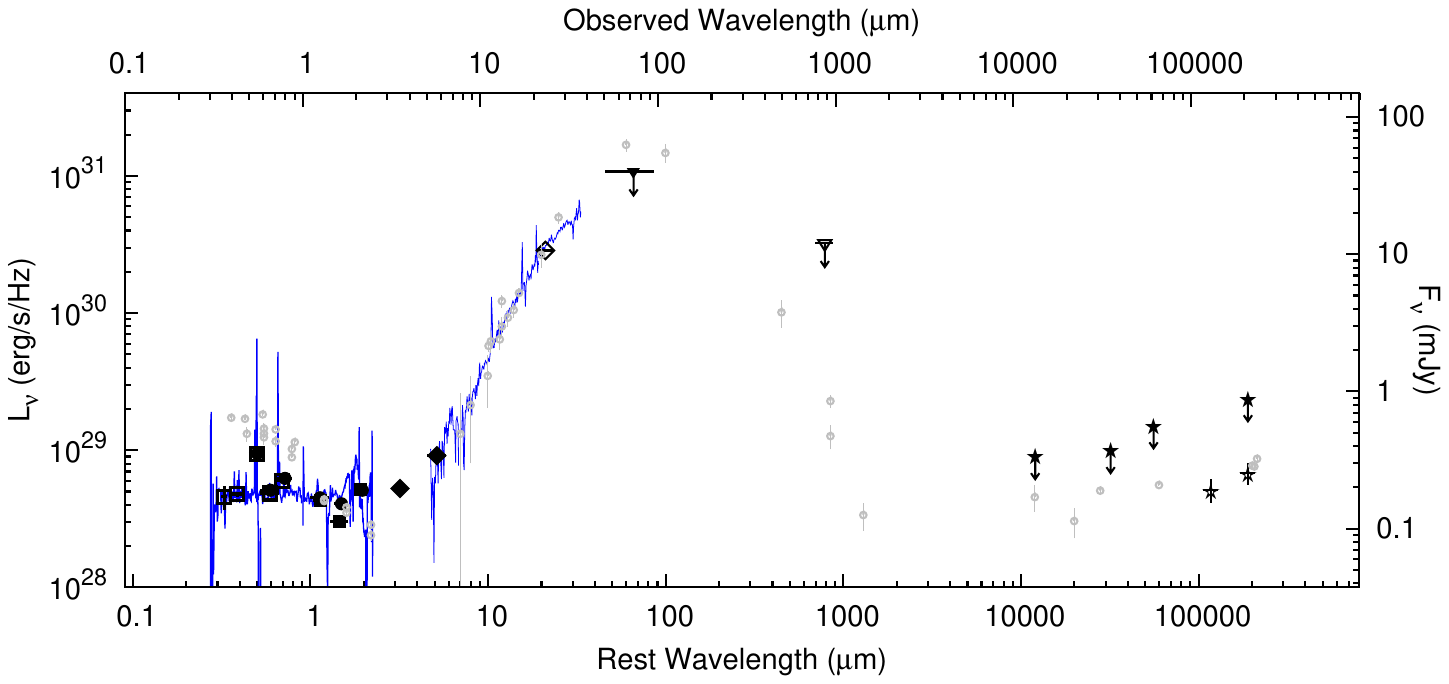}
 \caption{The SED of the host galaxy of GRB\,031203. The continuous lines
          (blue) are the X-shooter and \emph{Spitzer}-IRS spectra, open and
          closed squares are optical \citep{2007A&A...474..815M} and near-IR
          \citep{2004ApJ...611..200P} photometry respectively, while closed
          circles represent additional optical and near-IR photometry from
          \citet{2004ApJ...609L...5M} respectively.  Data from
          \emph{Spitzer}-IRAC (diamonds) and MIPS-$24\,\mu$m (open diamond)
          as well as ATCA (open stars) are also plotted.  Finally,
          upper limits are shown from MIPS-SED mode (filled, down
          triangles), APEX $870\,\mu$m (open, down triangles) and VLA
          \protect\citep[filled stars,][]{2004Natur.430..648S}.  The
          SED has been corrected for a Galactic extinction of E($B-V$)=1.04. 
          Marked in small open circles (grey), is the SED of the
          actively--star-forming blue compact dwarf galaxy II\,Zw\,40 which
          has a similar metallicity and radiation field.  Its SED is scaled
          by a factor of 17, the ratio of the [\ion{Ne}{3}]\,15.6\,$\mu$m
          emission lines. The best-fit SED template modelled with the GRASIL
          code is plotted as a dotted line.}
 \label{fig:SED}
\end{figure*}

The combination of 0.3--2.5\,$\mu$m spectroscopy from X-shooter and
5--35\,$\mu$m spectroscopy from the IRS, combined with limits in the FIR and
sub-mm and detections at radio wavelengths, allow us to get a fairly
complete picture of the properties of HG\,031203 via its SED
(Fig.~\ref{fig:SED}). 

\subsubsection{Foreground dust extinction}
The first outstanding question is the Galactic dust obscuration toward this
galaxy.  The Balmer decrement, which, due to the low redshift, essentially
gives us the total reddening along the line of sight, yields E$(B-V)=1.17$
\citep{2004ApJ...611..200P}: it is not entirely clear, however, what
fraction of this extinction is in the Galaxy and what fraction is in
HG\,031203.  Before discussing the foreground extinction, it should be noted
that our main results for HG\,031203 are insensitive to the precise value of
the extinction fraction in our Galaxy.  The mid-IR emission is largely
unaffected by this level of extinction.  Our primary age indicator for the
galaxy population, D4000, is also insensitive to the precise extinction
correction applied, as are the most important optical line ratios since they
lie very close in wavelength space (e.g.\ \ion{O}{3}/H$\beta$).
 
The dust maps of
\citet{1998ApJ...500..525S} yield E$(B-V)=1.04$ in the Galaxy alone in this
direction and this value has been used by several groups 
\citep{2004A&A...419L..21T,2004ApJ...609L..59G,2004ApJ...609L...5M,2004ApJ...608L..93C}. 
While these measurements seem to imply that the correction factor to obtain
the intrinsic galaxy spectrum is E$(B-V)=1.04$, \citet{2004ApJ...611..200P}
and \citet{2007A&A...474..815M} in fact argue that the value from
\citet{1998ApJ...500..525S} is overestimated along this sightline and
suggest the foreground Galactic extinction is E$(B-V)=0.7$--0.8, with
E$(B-V)\sim0.3$--0.4 residing in HG\,031203.  Recently,
\citet{2010arXiv1009.4933S} have indicated that the global normalisation for
E($B-V$) in \citet{1998ApJ...500..525S} may be overestimated by $\sim14$\%. 
There are, however, significant variations in the normalisation over the sky
($\sim20\%$): in particular, in this direction in longitude, the
normalisation is somewhat higher, but for high E($B-V$) sightlines, the
normalisation is somewhat lower.  Furthermore, for very high extinction
sightlines, the extinction-estimation method used is not effective.  This
leaves the situation somewhat confused.

Another line of attack to address
the level of the Galactic extinction is to use the Leiden-Argentine-Bonn
Galactic \ion{H}{1} maps, which indicate a column density of
$5.0\times10^{21}$\,cm$^{-2}$ in this direction \citep{2005A&A...440..775K}. 
The gas to dust correlation of \citet{1995A&A...293..889P} and more recently
\citet{2003A&A...408..581V}, suggest an E($B-V$) of 0.91 and 1.03
respectively, assuming standard ISM abundances; but the measurement is
abundance sensitive and may be lower by 20\% for a different choice of
metallicity \citep{2003A&A...408..581V}.  The situation is therefore still
complex.

However, in fitting the SED of HG\,031203, we find that the total
star-formation rate (SFR) is strongly inconsistent (larger than a factor of
two) with the H$\alpha$-derived SFR if we assume a low Galactic extinction
(E($B-V$)=0.78).  We require an extinction close to the value of
\citet{1998ApJ...500..525S} to reconcile the SED-derived SFR with the SFR
derived from H$\alpha$ by \citet{2004ApJ...611..200P} and
\citet{2007A&A...474..815M}.  However, the presence of a significant MIR
dust emission in HG\,031203 suggests that the UV light suffers some dust
extinction in the host, and our modelling of the full SED suggests A$_V$ of
a few tenths of magnitudes (see below).  This is consistent with the
observed Balmer decrement since it adds relatively little (E$(B-V)\sim0.1$)
to the overall reddening.  On the basis of these arguments, we simply use
E$(B-V)=1.04$, the value from the \citet{1998ApJ...500..525S} dust maps, to
correct for Galactic foreground extinction in our analysis of the HG\,031203
SED.

\subsubsection{Comparison to active blue compact dwarf galaxies}

The strong nebular emission lines, extreme hard line ratios, lack of PAH
emission features, and probable hot dust temperature, suggest a close
similarity between HG\,031203 and blue compact dwarf galaxies.  This
suggestion was made by \citet{2005NewA...11..103S} when analysing the
optical properties of low redshift SN-GRB host galaxies, including
HG\,031203.  In particular, HG\,031203 is similar to those classified as
being in an ``active'' mode of star-formation
\citep[e.g.][]{2005A&A...434..849H}.  The IR line ratios in fact, suggest
that HG\,031203 is among the more extreme BCDs in terms of the hardness of
its radiation field (Fig.~\ref{fig:line_ratios}).  A striking similarity
exists in the mid-IR emission of HG\,031203 and II\,Zw\,40, which has a very
similar metallicity and line ratios \citep{2008ApJ...673..193W} and a nearly
identical IRS spectrum in terms of slope and strength of features
\citep{2006ApJ...639..157W}.  We have plotted the SED of II\,Zw\,40 from
\citet{2005A&A...434..849H} on top of the SED of HG\,031203 in
Fig.~\ref{fig:SED} scaled by the ratio of the [\ion{Ne}{3}]\,$15.6\,\mu$m line
fluxes.  It is clear that it is only in the mid-IR that the spectral
similarity holds.  The UV/optical continuum of II\,Zw\,40 is considerably
bluer; its IR peak is brighter than is possible for HG\,031203, and is
likely somewhat cooler.  Both facts might be explained by a distribution of
dust in II\,Zw\,40 that is different, possibly distributed more uniformly,
and less densely located in star-forming regions than is the case in
HG\,031203.

\subsubsection{Modelling the SED}

We attempted to fully reproduce the UV--radio SED of HG\,031203 using the
GRASIL code \citep{1998ApJ...509..103S}.  The best-fitting SED model (see
Fig.~\ref{fig:SED}) was obtained with the fitting procedure outlined in
\citet{2010A&A...514A..67M} using the templates of
\citet{2007ApJ...670..279I} and \citet{2008ApJ...672..817M} with the PAH
emission removed and restricting the time of the total galaxy evolution to
$200$\,Myr.  This attempt failed to reproduce the SED on two points.  First
the observed 3.6\,$\mu$m flux was too high with respect to any of the
models.  This excess is well-known in dwarf galaxies and is speculated to be
due to a hot dust component, strong Br$\alpha$ emission, or nebular
continuum emission \citep{2009AJ....138..130S}.  Given the effect of the
very strong nebular emission lines on other photometric datapoints (e.g.\
the elevated $V$-band magnitude, Fig.~\ref{fig:SED}), line emission from
Br$\alpha$ may be sufficient to explain the observed 3.6\,$\mu$m excess
here.  The second point of failure relates to the 4000\,\AA\ break, where
none of the templates could reproduce the lack of a break in the spectrum
HG\,031203, in spite of the restriction to younger stellar templates.

Finally, the dust mass is not very well-constrained with this dataset, since
we do not quite cover the peak of the thermal emission component.  An
indication of a turnover in the IRS spectra at the longest wavelengths
(Fig.~\ref{fig:SED}) suggests the peak dust temperature is very high, as
observed in dwarf galaxies \citep{2009A&A...508..645G}, perhaps hotter even
than the 45\,K suggested for sub-mm--detected GRB hosts
\citep{2008ApJ...672..817M}.  However our constraints at longer wavelengths
do not allow us to exclude a large cool dust component.

%
%

\section{Discussion}\label{discussion}

We have found that HG\,031203 is an extreme star-forming galaxy, similar to,
but considerably more luminous than, nearby blue compact dwarf galaxies.  A
very hard radiation field is found from both optical and mid-infrared line
ratios, and is consistent with a very young star-forming population.  In
spite of its blue optical continuum, approximately as much energy is
reradiated by dust as comes directly from stars. This suggests a strongly
non-uniform dust distribution in the galaxy.

The host galaxy of GRB\,031203 is among the
brighter GRB hosts known because of its low redshift and intrinsic
luminosity.  Despite a high foreground extinction, it was already
well-characterised in the optical, showing a high star-formation rate and
relatively low metallicity.  \citet{2007A&A...474..815M} suggest it is
somewhat atypical of the few nearby GRB hosts so far observed in that it
lies, at low-metallicity and relatively high luminosity, off the
luminosity-metallicity locus of KPNO International Spectroscopic Survey
galaxies.  The deviation might be explained simply by the low pre-existing
stellar mass in this galaxy.  According to the fundamental plane for
star-forming galaxies of \citet{2010MNRAS.tmp.1314M,2010arXiv1005.0509L},
HG\,031203 should have a stellar mass of approximately
$\log({M_*/M_\odot})\sim9.5$, similar to the value obtained by
\citet{2010ApJ...721.1919C}.

Initial samples of gamma-ray burst host galaxies suggest they are typically
sub-luminous and blue \citep{2003A&A...400..499L} with young ages
\citep{2004A&A...425..913C}, high specific star-formation rates
\citep{2006ApJ...653L..85C,2009ApJ...691..182S,2010MNRAS.405...57S,2002ApJ...566..229C}
and irregular morphologies \citep{2006Natur.441..463F}.  All of these
conclusions however have been based on rather incomplete and arbitrarily
selected galaxy samples, affected by strong optical/UV selection bias
because of the requirement to have a detection of an optical afterglow to
localise the burst to a high enough accuracy to identify the host.  On the
other hand, analyses of the hosts of bursts believed to be dust-obscured are
consistent with this picture---sub-mm and radio observations yield only a
small fraction of detections of these targeted bursts
\citep{2004MNRAS.352.1073T,2003ApJ...588...99B} and star-formation limits
found in the radio \citep{2008ApJ...672..817M} and X-ray
\citep{2004A&A...425L..33W} show no evidence of very high star-formation
rates---suggesting that even the hosts of obscured bursts are not typically
ultraluminous IR galaxies (ULIRGs).  In particular, the hosts of GRBs
at low redshifts are consistent with this general outline
\citep{2005NewA...11..103S,2009ApJ...693..347M,2008ApJ...676.1151T,2007A&A...464..529W},
which means that despite the strong evolution of star-forming galaxies from
$z\gtrsim3$ to $z=0$, GRBs offer a good way of targeting closer
sites of very young star-formation that are typical of galaxies in the early
universe \citep[e.g.][]{2009ApJ...702..377K}.

\subsection{HG\,031203 as a high-redshift star-forming galaxy}
HG\,031203 may offer a good low-redshift archetype of high-redshift
star-forming galaxies.  Its extremely young age, blue optical continuum, and
low stellar mass bears a striking resemblance to models of $z$ and $Y$-band
drop-outs detected with \emph{HST}.  A $z\sim2$ analogue of these galaxies
was suggested recently by \citet{2010ApJ...719.1168E}, and it is interesting
to note that HG\,031203 has emission line ratios and estimated stellar mass
very similar to that galaxy.  Furthermore, the fact that HG\,031203 was
selected by the presence of a GRB---known to be associated with the deaths
of massive stars, and now among the most distant spectroscopically-confirmed
objects---adds to the interest with which HG\,031203 must be held as a
nearby example of the high-redshift mode of massive star-formation.

Our data on HG\,031203 bear on the current debate surrounding the SEDs of
$z\sim7$ and $z\sim8$ galaxies detected via $z$ and $Y$-band drop-outs.  It
has been suggested that the elevated fluxes (above the extrapolation of the
restframe UV continuum) redward of the 4000\,\AA\ break in these objects as
observed by IRAC in stacked 3.6 and $4.5\,\mu$m images, are indicative of a
substantial 4000\,\AA/Balmer break.  From these detections a stellar age of
$\sim300$\,Myr is inferred for these galaxies
\citep{2010ApJ...708L..26L,2010ApJ...716L.103L,2010arXiv1006.4360B}. 
However it has been pointed out, and demonstrated via stellar synthesis
models, that nebular line emission, in particular very strong [\ion{O}{3}]
lines, can produce an elevation in the IRAC bands similar to that observed
\citep{2010A&A...515A..73S,2010MNRAS.402.1580O}.  Using models containing
nebular line emission, a population as young as 5\,Myr provides a reasonable
fit to the data \citep{2010A&A...515A..73S}.  The question of old stellar
populations in galaxies at $z\sim7-8$ is critically important, as it has a
strong bearing on when the first population of stars formed and whether
these stars formed at a time that could have allowed them to cause
reionisation.  While it has been argued on a number of grounds that the
elevated IRAC fluxes are indeed related to an old stellar population in
these very early galaxies \citep{2010arXiv1006.4360B}, the arguments are
circumstantial and the subject is still open.

\begin{figure}
 \includegraphics[angle=0,width=\columnwidth,viewport=160 59 660 554,clip=]{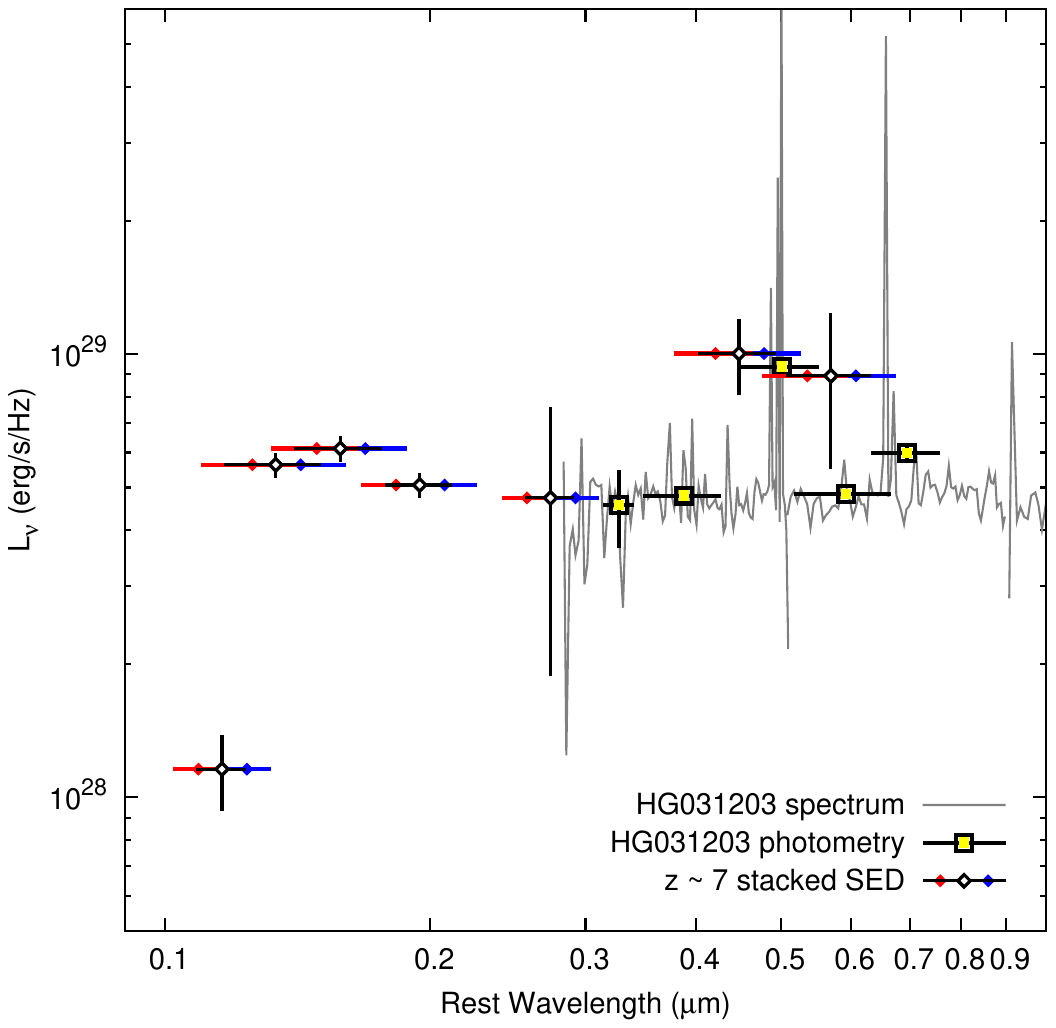}
 \caption{The UV-optical SED of the host galaxy of GRB\,031203 (grey solid
          line and black open squares). The $V$-band photometry is clearly
          strongly elevated simply due to the [\ion{O}{3}] emission lines. 
          Overplotted (diamonds) is the stacked SED of faint $z\sim7$, $z$-band
          dropouts from \protect\citet{2010ApJ...708L..26L}
          scaled by 0.36 to match the luminosity of HG\,031203.  The blue and
          red diamonds represent redshifts of 6.4 and 7.4 respectively,
          indicating the approximate range of rest wavelengths entering the
          stacked filters.  The elevation of photometry near the
          [\ion{O}{3}] lines and the slope of the UV-optical continuum are
          similar in both SEDs suggesting the galaxies have similar dominant
          stellar ages, $\lesssim10$\,Myr.}
\label{fig:labbe_SED}
\end{figure}

In Fig.~\ref{fig:labbe_SED} we overplot the SED from
\citet{2010ApJ...708L..26L} of faint $z\sim7$ drop-out galaxies (scaled by
0.36), on the SED of HG\,031203.  The approximate similarity in UV slope
(though there is some uncertainty here due to the high foreground Galactic
dust column), in the total luminosity, and in the relative elevation of the
broadband fluxes, show clearly that HG\,031203 is indeed a reasonable
analogue of these $z\sim7$ candidates.  Furthermore, we see in this analogue
precisely the broadband photometry behaviour suggested by the modelling of
\citet{2010A&A...515A..73S}.  The effect only occurs because of the stacking
of sources.  At $z=7$, the line lies in neither IRAC band, but the stacking
of sources in the range $z=6.4-7.4$ and the relatively rectangular shape of
the bandpasses raises both IRAC bands.

Lastly, in spite of the very young stellar age indicated by the lack of a
4000\,\AA\ break, HG\,031203 is clearly dusty, with as much emission from
star-formation emerging in the mid-IR via dust-reprocessing as there is
emerging in the UV.  The best-fit GRASIL model also suggests that the UV
emission is still significantly obscured ($A_V\sim 0.3$), although it is
obvious that this must be the case from the mid-IR emission.  This latter
point also echoes the claim from \citet{2010A&A...515A..73S} based on
modelling, that drop-out galaxies at $z\sim7$ may still be substantially
obscured in spite of their blue continua.  From sub-mm and mm observations,
we are already aware of the presence of large quantities of dust at $z\sim6$
in QSOs.  However, the possible presence of dust in significant quantities
in very young galaxies at $z\sim7$, suggests that dust is formed very
rapidly.  Typically, very rapidly is taken to mean a few hundred million
years \citep[e.g.][]{2007ApJ...662..927D,2009MNRAS.397.1661V}.  However, if
the dust is formed in association with the formation of stars, whether in
SNe
\citep{2007ApJ...662..927D,2003Natur.424..285D,2009MNRAS.394.1307D}, or
grown in the ISM
\citep{2009ASPC..414..453D,2010ApJ...712..942M,2010arXiv1006.5466M} the
timescales could in principle be as short as several million years.  This
has major implications for the detection of galaxies of this luminosity with
NIRCAM, NIRSPEC and MIRI onboard \emph{JWST} at $z\sim7$--10.

%
%
\section{Conclusions\label{conclusions}}

We have presented the first mid-IR spectrum of a GRB host galaxy as
well as the first X-shooter spectrum of a GRB host.  These spectra show strong
nebular lines associated with recent star-formation, but no PAH features. 
We conclude that HG\,031203 is an extreme star-forming galaxy with a
radiation field among the hardest known for such a galaxy, in both its
obscured and unobscured regions.  The broadband SED of HG\,031203 has a
moderate luminosity dust emission peak, likely with a high temperature,
similar to local blue compact dwarf galaxies.  If dust can be formed in
association with supernovae, it is possible that such a dust peak may exist
even in very young galaxies in the early universe, making such galaxies
potentially much easier to detect with \emph{JWST}.  The lack of a
4000\,\AA\ break and the high-ionisation emission line ratios in the MIR
spectrum of HG\,031203 suggests a dominant stellar age $\lesssim 10$\,Myr. 
HG\,031203 has a UV-optical SED similar to $z$-band dropout galaxies at
$z\sim7$.  This and its very young age indicate that this GRB host galaxy at
$z\sim0.1$ may be an excellent analogue of the population of star-forming
galaxies at very high redshifts currently being discovered by \emph{HST}.

\begin{acknowledgements} 
The Dark Cosmology Centre is funded by the DNRF.  JF acknowledges support
from Instrument Center for Danish Astrophysics.  We would like to thank
S.~Toft and J.~Richard for discussions and comments on the manuscript.  We
are also grateful to the ESO commissioning team, in particular S.~D'Odorico,
J.~Vernet, H.~Dekker, J.~Lizon, R.~Castillo, M.~Downing, G.~Finger,
G.~Fischer, C.~Lucuix, P.~DiMarcantonio, E.~Mason, A.~Modigliani, S.~Ramsay
and P.~Santin.  The Australia Telescope Compact Array is part of the
Australia Telescope which is funded by the Commonwealth of Australia for
operation as a National Facility managed by CSIRO.  This publication is
based on data acquired with the Atacama Pathfinder Experiment (APEX).  APEX
is a collaboration between the Max-Planck-Institut fur Radioastronomie, the
European Southern Observatory, and the Onsala Space Observatory. 
\end{acknowledgements}

\bibliography{mnemonic,grbs}

\end{document}